# Terahertz Imaging System with On-Chip Superconducting Josephson Plasma Emitters for Nondestructive Testing

Manabu Tsujimoto[1*†], Kaveh Delfanazari[2*†], Takanari Kashiwagi[3], Toshiaki Hattori[3], and Kazuo Kadowaki[3]

1. Research Center for Emerging Computing Technologies, National Institute of Advanced Industrial Science and Technology (AIST), Central 2, 1-1-1 Umezono, Tsukuba, Ibaraki 305-8568, Japan
2. Electronics and Nanoscale Engineering Division, James Watt School of Engineering, University of Glasgow, Glasgow G12 8QQ, United Kingdom
3. Faculty of Pure and Applied Sciences, University of Tsukuba, 1-1-1 Tennodai, Tsukuba, Ibaraki 305-8573, Japan

**Abstract**

Compared with adjacent microwaves and infrared frequencies, terahertz (THz) frequency offers numerous advantages for imaging applications. The unique THz spectral signatures of chemicals allow the development of THz imaging systems for nondestructive tests and the evaluation of biological objects, materials, components, circuits, and systems, which are especially useful in the security, medical, material, pharmaceutical, aeronautical, and electronics industries. However, technological advancements have been hindered owing to the lack of power-efficient and compact THz sources. Here, we use high-temperature superconducting monolithic sources known as Josephson plasma emitters (JPEs)—which are compact, chip-integrated coherent and monochromatic sources of broadly tunable THz waves—and report the art of non-destructive imaging of concealed metallic surgical blades, floppy disks, dandelion leaves, and slices of pork meat in the THz spectral range. The quality of the images, exhibiting high-contrast differentiation between metallic and non-metallic parts, making different features of objects visible, and targeting different powders, demonstrates the viability of this THz imaging system for nondestructive, contactless, quick, and accurate environmental monitoring, security, medicine, materials, and quantum science and technology applications.

## Introduction

In recent years, there has been a growing interest in terahertz (THz) imaging owing to its unique capabilities for the nondestructive and non-invasive inspection of various materials and structures [1–3]. THz waves, with frequencies between 0.1 and 10 THz, can penetrate through various materials, such as plastics, fabrics, and ceramics, while also revealing their internal structures and chemical compositions [4]. THz imaging has diverse applications in fields such as security, medicine, and the materials sciences. For example, in security applications, THz imaging has been used for the detection of concealed weapons, explosives, and drugs, whereas in medicine, it has been utilized for the

---

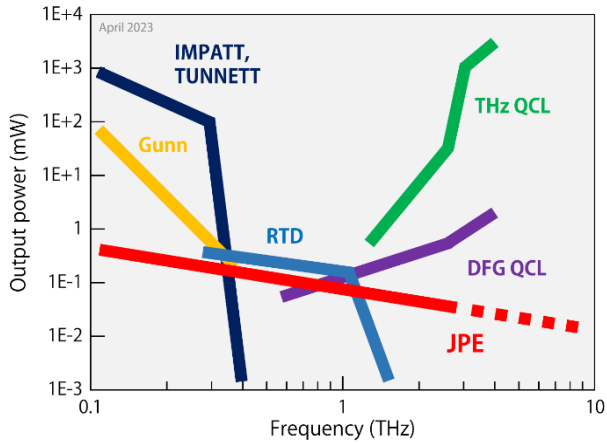

Fig. 1: Current status of solid-state on-chip THz sources, including IMPATT diodes, TUNNETT diodes, THz QCL, DFG QCL, RTD, and JPE. The output power is plotted as a function of generated frequency.

detection of skin cancer, dental diagnoses, and monitoring wound healing [5].

The two major sources of THz imaging systems are continuous-wave and time-domain spectroscopy (THz-TDS). Continuous-wave sources include quantum cascade lasers (QCLs), backward-wave oscillators, and optically pumped THz lasers [6]. Significant progress has been made in THz pulsed imaging based on THz-TDS [7]. However, these imaging systems have slow scanning rates, low power, and bulky THz sources; thus, there is an urgent need for efficient, compact, and tunable radiation sources. Figure 1 illustrates the latest advances in solid-state on-chip THz sources, including impact-ionization avalanche transit-time (IMPATT) diodes, tunneling transit-time (TUNNETT) diodes, QCLs, difference-frequency-generation QCLs (DFG QCLs), and resonant-tunneling diodes (RTDs). The output power is plotted as a function of the generated frequency.

In 2007, we discovered a remarkable phenomenon in which a high-transition temperature (high-$T_c$) superconducting $Bi_2Sr_2CaCu_2O_{8+\delta}$ (BSCCO) compound generated intense and coherent electromagnetic waves in the THz frequency region associated with Josephson plasma excitation [8–12]. This compound is known as an intrinsic Josephson junction (IJJ) system, in which insulating $Bi_2O_2$ planes are sandwiched between $CuO_2$ double planes responsible for superconductivity [13]. THz sources based on Josephson plasma oscillations in IJJs are referred to as Josephson plasma emitters (JPEs). Studies have shown that the JPEs can emit continuous and microwatt power THz waves at tunable frequencies between 0.15–11 THz [14–17] with a spectral linewidth of down to 23 MHz [18] and a maximum operating temperature of above 77 K [19,20]. Solid-state JPEs are extremely small, and their output power is sufficiently stable during operation [21]. Therefore, JPEs with high spectral power output, narrow linewidth, and tunable frequency range are promising candidates for THz imaging and camera technology development [22,23].

We developed THz imaging systems for practical use based on JPE as the source, using both transmission and reflection modes [24–27]. Furthermore, the small size of the JPE THz camera, owing to the micrometer-sized superconducting cavity, facilitates the integration of cryogenic THz camera circuitry for imaging applications in quantum computing and processing, such as the imaging of superconducting microwave resonators [28,29].

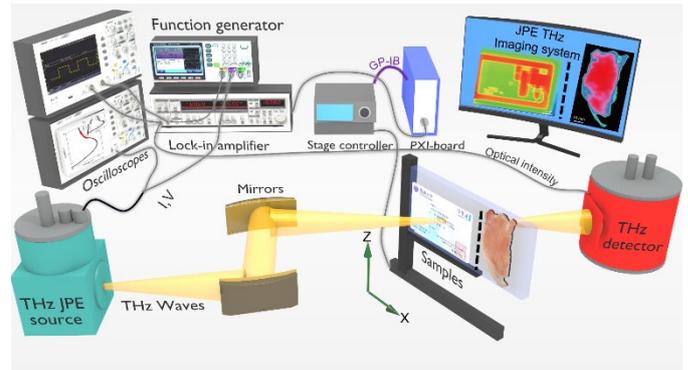

Fig. 2: Schematic of a THz imaging system with on-chip JPE sources. A micron-sized JPE source is integrated on a chip and mounted in the cold finger of a He-Flow cryostat. THz waves generated by the JPE source are focused on samples using off-axis parabolic mirrors. Samples are scanned on a two-dimensional stage and lock-in amplifier is used to reduce background noise. Real-time imaging is obtained by measuring sample position and THz detector output.

This study aims to evaluate the applicability of a JPE-based THz camera imaging system for the practical use of security in observing concealed objects, industrial inspection in agriculture and food, and nondestructive testing of integrated circuits, biological tissues, and nanomaterials. We present the THz images of four samples: surgical blades within a paper envelope, floppy disk, Dandelion leaf, and slice of pork. In addition, we employed an imaging system to accurately measure the absorptance of the powder samples.

### Experimental

A schematic of the JPE THz imaging system is shown in Fig. 2. The JPE source was integrated on a chip and assembled in the He-flow cryostat of a transmission-type imaging system [24]. The system comprised two off-axis parabolic mirrors with focal lengths of 152.4 mm and 220 mm and diameters of 75 mm. These mirrors were used to focus the THz waves onto the sample position. The imaging object was mounted on a two-dimensional scanner (SIGMA KOKI Co., SGAMH26-200) and scanned in the X (horizontal) and Z (vertical) directions at speeds of up to 80 mm/s, corresponding to a 5-ms time constant per data point. The measurement step was set to a submillimeter, which is equivalent to the wavelength. Although the JPE source can be modulated relatively faster, the minimum time constant of the THz detector (InSb hot-electron bolometer) limits the maximum scan speed. However, the imaging was performed at 32 mm/s to obtain sufficient data points. An electrically modulated THz wave was generated by applying a low-amplitude square wave at 10 kHz to a constant DC offset voltage using a function generator (Hewlett-Packard Co., 33120A). To avoid far-infrared background noise caused by ambient spurious radiation, we employed a lock-in detection technique, which is superior to optical chopping and other methods. We adjusted the DC offset level to maximize the intensity of the generated wave while minimizing the power fluctuations by adjusting the amplitude of the square wave. Imaging data were acquired by simultaneously measuring the sample positions (X, Z) and bolometer output voltage, which were acquired using an AD converter in the PXI bus system (NI Co.). The LabVIEW program was used to set the scanning parameters, including speed, step, and measurement area. Data were stored directly on a hard disk drive and displayed in real-time on a PC screen.

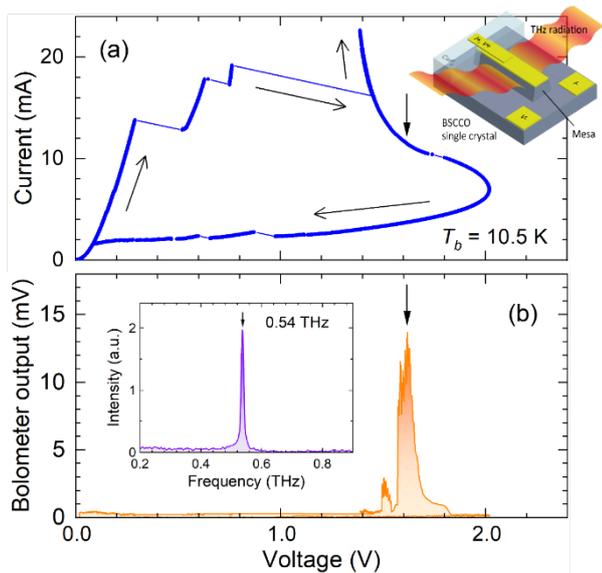

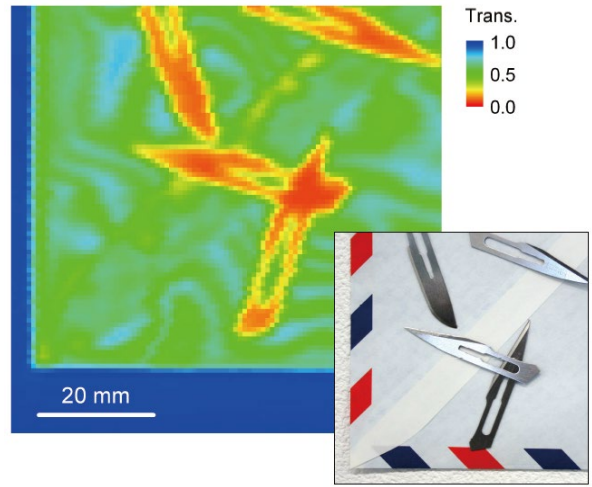

Fig. 3: (a) IVCs and (b) the THz detector output of the high-$T_c$ superconducting JPE measured at $T_b$ = 10.5 K. The inset in (a) presents a schematic of the JPE source. The inset in (b) displays the emission spectrum at 11.5 mA, which was measured using an FT-IR spectrometer.

Fig. 4: (a) THz image of surgical blades enclosed in a paper envelope, with a fixed emission frequency of $f$ = 0.54 THz. The image displays the THz wave's transmittance in a color plot. An optical image is provided alongside the THz images. The metallic blades show no transmittance, and the transmittance of the paper envelope is 79%.

A schematic of the on-chip JPE THz source used in this study is shown in the inset of Fig. 3(a). The fabrication process is described in detail in a previous study [30]. The JPE source used in this study was identical to that used in our previous study [24], and its characteristics, outlined below, were the same as those reported in the literature. The device cavity (mesa) dimensions of 62×400×1.9 μm³ were measured with a stylus profiler. The current-voltage characteristics (IVC) measured at a bath temperature of $T_b$ = 10.5 K are shown in Fig. 3(a). IVCs were obtained using cyclic DC scans. The finite voltage observed below $I$ = 13.8 mA, where all stacked IJJs are in the superconducting state, is attributed to the contact resistance at the interface between the BSCCO and gold films used as electrodes.

The bolometer output monitored during the IVC measurements is shown in Fig. 3(b). Two distinct peaks were observed in the bolometer output: one with a lower intensity at a higher current, and the other with a higher intensity at a lower current. When calibrated using the sensitivity of the detector at $V$ = 1.618 V and $I$ = 11.5 mA, the maximum output of 13.7 mV corresponds to an integrated radiation power of 0.2 μW when calibrated using the sensitivity of the detector. However, note that the total output power from the device must be evaluated separately by measuring the spatial radiation patterns. The radiation spectrum measured at 11.5 mA using an FT-IR spectrometer (JASCO Co., FARIS-1) is presented in the inset of Fig. 3(b), where a sharp emission peak at $f$ = 0.54 THz is observed. Using the AC Josephson relation with a contact resistance of 13 Ω, the number of resistive IJJs can be estimated to be $N$ = 1320, which is consistent with a JPE mesa height of 1.9 μm.

## Results and discussion

This study demonstrates nondestructive THz frequency imaging of various objects, including surgical blades within a paper envelope, floppy disk, dandelion leaf, and slice of pork meat, using JPEs with a fixed radiation frequency of $f$ = 0.54 THz. The radiation frequency can be varied by adjusting the bias voltage $V$ based on the Josephson relation. The emission frequency was consistent with the geometrical cavity resonance condition [31–36] and was explained using a microstrip antenna model [37,38]. The acquisition time for each transmission image was approximately 15 min, and the stability of the JPE radiation power was verified by observing a constant bolometer output in the blue area in Fig. 4, where no absorbing object was present. The signal-to-noise ratio was estimated to be approximately 130 because the noise level was found to be below 0.1 mV. The estimated spatial resolution of our imaging system was approximately 1 mm, which is comparable to the diffraction limits of optical systems. The diffraction-limited spot size is given by $s \approx \lambda \cdot F$ = 1.13 mm, where $F$ = 2.0 represents the off-axis parabolic mirror's $f$-number. Some studies have reported spatial resolutions as low as $0.15\lambda$ [39].

To evaluate the viability of finding concealed objects, the THz imaging of surgical blades within a paper envelope is demonstrated in Fig. 4. The transmittance of a single sheet of paper was calculated to be 79%, whereas the metallic objects did not exhibit transparency. In the green area, several interference fringe-like patterns are visible, which are recognized as iso-thickness interference patterns caused by monochromatic THz waves. The incident THz wave had a wavelength of $\lambda = c_0/f$ = 560 μm and tended to get multiply reflected at the inner walls of the paper envelope and interfere with itself. The interference condition can be expressed as $2d_m = (\lambda/2) \times 2m$, where $d_m$ is the interspace distance of the inner walls and $m$ is an integer. Therefore, the distance between neighboring fringes can be used to accurately measure the thickness of the paper and/or the interspace of the envelope; this technique, known as THz interferometric imaging, uses coherent THz pulses [40,41].

A THz image of a floppy disk is shown in Fig. 5. Various features of the plastic cover are distinguishable, as each causes a distinct transmittance of the THz wave. The metal hub and shutter with their apertures showed no transmittance. The transmittance of the circular magnetic recording disk was approximately 50%, and that of the hollow lower part of the disk was approximately 100%.

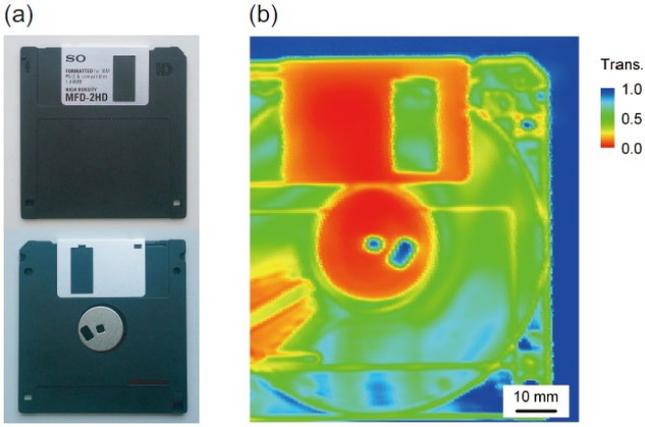

Fig. 5: (a) Optical and (b) THz images of a floppy disk, with a fixed emission frequency of $f$ = 0.54 THz. The metallic shutter and hub are opaque. The circular recording disk is half transparent and various features of the plastic disk and buried components are distinguishable.

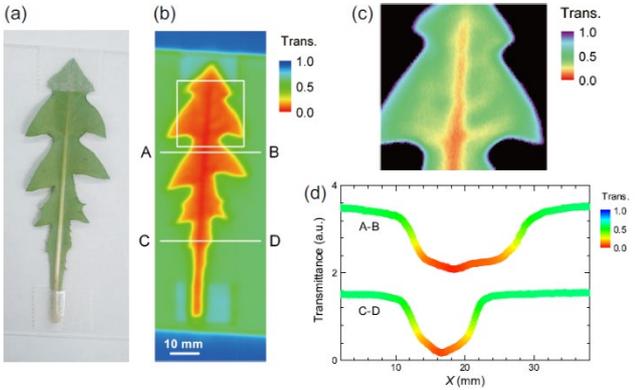

Fig. 6: Nondestructive THz image of a dandelion leaf. (a) Optical and (b) THz transmission image, as well as (c) magnified image of the area indicated by the white line in (b). (d) Line profile results for positions A-B and C-D in (b).

THz imaging has been shown to be useful for inspecting and monitoring agricultural products [42]. Fig. 6 presents a THz image of a dandelion leaf. The image shows the leaf thorns and adhesive tape used to attach the leaf to the sample holder. Furthermore, the main and side veins of the leaf are visible in Fig. 6(c), which shows a magnified image of the area indicated by the white rectangle in Fig. 6(b). The transmission along the white lines A-B and C-D is shown in Fig. 6(d), where the drop in transmission in the middle allowed for discrimination of the main vein in the leaf.

THz imaging can distinguish between normal and cancerous tissues based on differences in their water content [43]. In addition, the freshness and quality of meat can be determined by measuring the reduction in the water content of bad meat [44]. The image of a slice of pork in Fig. 7 shows the regions of protein and fat. The lean meat in the middle appeared opaque, whereas the fat around the slice appeared transparent, possibly owing to differences in thickness. Therefore, THz imaging can be used to detect the proportions of fat and lean meat.

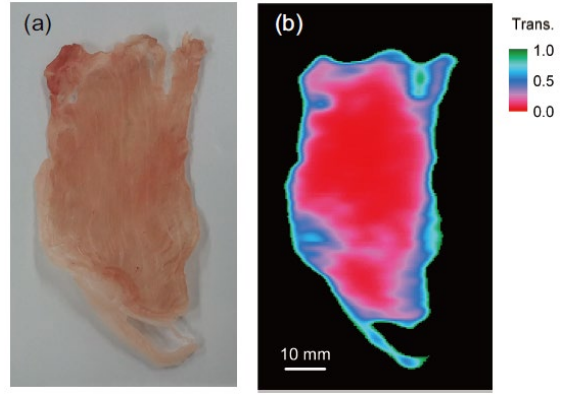

Fig. 7: (a) Optical and (b) THz images of a slice of pork meat with a fixed emission frequency of $f$ = 0.54 THz. The middle lean meat's transmittance is approximately zero while the fat meat is almost transparent.

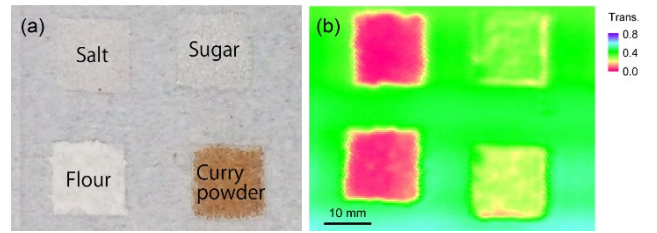

Fig. 8: (a) Optical and (b) THz images of four powder samples measured at $f$ = 0.54 THz. The type of powder sample is noted on the optical image. Different transmittance is the result of different densities, intramolecular vibrations, and hydrogen bonds in different powders.

Table I: Absorptance $A_p$ of eight powder samples.

| Powder | $A_p$ (%) |
|---|---|
| Salt | 82 |
| Sugar | 10 |
| Pepper | 72 |
| Flour | 82 |
| Curry | 41 |
| Instant soup | < 1 |
| Coffee | 9 |
| Stomach medicine | 86 |

Absorption experiments were performed on multiple powder samples using a transmission imaging system. Fig. 8 displays the optical and THz images of four powders: salt (top left), sugar (top right), flour (bottom left), and curry powder (bottom right). Each powder was sprinkled in small quantities on a 1 mm-thick quartz plate and secured with 15×15 mm² double-sided tape. The quartz plate was scanned in both the X- and Z-directions at a speed of 32 mm/s using a two-dimensional scanner.

The THz absorptance $A_p$ of each powder sample was easily determined by subtracting the contribution of absorption from the quartz plate and double-sided tape. Table I summarizes the $A_p$ values of the eight powdered samples. Notably, common samples, such as salt and sugar, exhibited a wide range of $A_p$ values despite their similar appearance. This suggests that the intramolecular vibrations and hydrogen bonds in the powder samples are highly sensitive to THz waves. The powder type can be determined by its absorptance. The sample thickness $t_p$ can be calculated as $t_p = W/\rho S$, where $W$ and $\rho$ denote the weight and density of the powder, respectively, bonded by the double-sided tape with the area $S$. The absorption coefficient $\alpha_p$, a material-specific physical property can be obtained by $\alpha_p = -t_p^{-1} \ln(1 - A_p)$.

Our imaging system can quickly obtain $\alpha_p$ values in the THz frequency range, enabling the identification of specific types and amounts of materials in sealed envelopes without being exposed to the atmosphere. Although previous studies explored multispectral THz imaging and powder detection in envelopes [45,46], they did not leverage the spectral fingerprints of the target material. Practical applications of this technique should consider environmental factors, such as temperature and humidity, as THz waves are readily absorbed by atmospheric moisture.

## Conclusion

We demonstrated a nondestructive THz imaging system that utilizes high-$T_c$ superconducting JPEs as coherent THz sources. The JPEs offered intense, coherent, and monochromatic THz radiation capable of tuning the THz waves. To assess the practical potential of JPE-based THz imaging systems, we conducted imaging experiments at a frequency of 0.54 THz on various objects, including surgical blades inside a paper envelope, floppy disk, dandelion leaf, and slice of pork meat. In addition, we utilized an imaging system to measure the absorptance of various powder samples. The unique properties of superconducting JPEs make THz JPE-based imaging systems a promising approach for nondestructive testing, evaluation, inspection, imaging, and processing in a wide range of applications from fields such as security, medicine, and material science to quantum computing, sensing, and communication systems.

## Acknowledgment

The authors express their gratitude to M. Sawamura, R. Nakayama, and T. Kitamura for their technical assistance. The insightful discussions with H. Minami were invaluable. The single crystals utilized in this study were grown by T. Yamamoto at the University of Tsukuba. The authors acknowledge funding support from the CREST-JST (Japan Science and Technology Agency) and the World Premier International Research Center Initiative (WPI)-MANA (Materials Nanoarchitectonics) project (NIMS). K. Delfanazari also acknowledges the Japanese Government's Monbukagakusho Scholarship, and S. Kalhor and M. Zhang for their assistance.

## References

[1] K. Kawase, Y. Ogawa, Y. Watanabe, and H. Inoue, *Non-Destructive Terahertz Imaging of Illicit Drugs Using Spectral Fingerprints.*, Opt Express **11**, 2549 (2003).

[2] J. F. Federici, B. Schulkin, F. Huang, D. Gary, R. Barat, F. Oliveira, and D. Zimdars, *THz Imaging and Sensing for Security Applications—Explosives, Weapons and Drugs*, Semicond Sci Technol **20**, S266 (2005).

[3] D. M. Mittleman, *Twenty Years of Terahertz Imaging [Invited]*, Opt Express **26**, 9417 (2018).

[4] A. Redo-Sanchez, N. Laman, B. Schulkin, and T. Tongue, *Review of Terahertz Technology Readiness Assessment and Applications*, J Infrared Millim Terahertz Waves **34**, 500 (2013).

[5] X. C. Zhang, A. Shkurinov, and Y. Zhang, *Extreme Terahertz Science*, Nat Photonics **11**, 16 (2017).

[6] S. G. Pavlov, H.-W. Hübers, H. Riemann, R. K. Zhukavin, E. E. Orlova, and V. N. Shastin, *Terahertz Optically Pumped Si:Sb Laser*, J Appl Phys **92**, 5632 (2002).

[7] Q. Wang, L. Xie, and Y. Ying, *Overview of Imaging Methods Based on Terahertz Time-Domain Spectroscopy*, Appl Spectrosc Rev **57**, 249 (2022).

[8] L. Ozyuzer et al., *Emission of Coherent THz Radiation from Superconductors*, Science **318**, 1291 (2007).

[9] U. Welp, K. Kadowaki, and R. Kleiner, *Superconducting Emitters of THz Radiation*, Nat Photonics **7**, 702 (2013).

[10] I. Kakeya and H. Wang, *Terahertz-Wave Emission from Bi2212 Intrinsic Josephson Junctions: A Review on Recent Progress*, Supercond Sci Technol **29**, 073001 (2016).

[11] T. Kashiwagi et al., *The Present Status of High-$T_c$ Superconducting Terahertz Emitters*, Supercond Sci Technol **30**, 074008 (2017).

[12] K. Delfanazari, R. A. Klemm, H. J. Joyce, D. A. Ritchie, and K. Kadowaki, *Integrated, Portable, Tunable, and Coherent Terahertz Sources and Sensitive Detectors Based on Layered Superconductors*, Proceedings of the IEEE **108**, 721 (2020).

[13] R. Kleiner, F. Steinmeyer, G. Kunkel, and P. Müller, *Intrinsic Josephson Effects in $Bi_2Sr_2CaCu_2O_8$ Single Crystals*, Phys Rev Lett **68**, 2394 (1992).

[14] M. Tsujimoto et al., *Broadly Tunable Sub-Terahertz Emission from Internal Current-Voltage Characteristic Branches Generated from $Bi_2Sr_2CaCu_2O_{8+\delta}$*, Phys Rev Lett **108**, 107006 (2012).

[15] T. Kashiwagi et al., *A High-$T_c$ Intrinsic Josephson Junction Emitter Tunable from 0.5 to 2.4 Terahertz*, Appl Phys Lett **107**, 082601 (2015).

[16] E. A. Borodianskyi and V. M. Krasnov, *Josephson Emission with Frequency Span 1–11 THz from Small $Bi_2Sr_2CaCu_2O_{8+\delta}$ Mesa Structures*, Nat Commun **8**, 1742 (2017).

[17] H. Sun et al., *Compact High-$T_c$ Superconducting Terahertz Emitter with Tunable Frequency from 0.15 to 1 THz*, Applied Sciences **13**, 3469 (2023).

[18] M. Li et al., *Linewidth Dependence of Coherent Terahertz Emission from $Bi_2Sr_2CaCu_2O_8$ Intrinsic Josephson Junction Stacks in the Hot-Spot Regime*, Phys Rev B Condens Matter Mater Phys **86**, 060505 (2012).

[19] L. Y. Hao et al., *Compact Superconducting Terahertz Source Operating in Liquid Nitrogen*, Phys Rev Appl **3**, 024006 (2015).

[20] K. J. Kihlstrom et al., *Powerful Terahertz Emission from a $Bi_2Sr_2CaCu_2O_{8+\delta}$ Mesa Operating Above 77 K*, Phys Rev Appl **19**, 034055 (2023).

[21] H. Minami, I. Kakeya, H. Yamaguchi, T. Yamamoto, and K. Kadowaki, *Characteristics of Terahertz Radiation Emitted from the Intrinsic Josephson Junctions in High-$T_c$ Superconductor $Bi_2Sr_2CaCu_2O_{8+\delta}$*,

[22] K. Nakade, T. Kashiwagi, Y. Saiwai, H. Minami, T. Yamamoto, R. A. Klemm, and K. Kadowaki, *Applications Using High-$T_c$ Superconducting Terahertz Emitters*, Sci Rep **6**, 23178 (2016).

[23] R. Kleiner and H. Wang, *Terahertz Emission from $Bi_2Sr_2CaCu_2O_{8+x}$ Intrinsic Josephson Junction Stacks*, J Appl Phys **126**, 171101 (2019).

[24] M. Tsujimoto, H. Minami, K. Delfanazari, M. Sawamura, R. Nakayama, T. Kitamura, T. Yamamoto, T. Kashiwagi, T. Hattori, and K. Kadowaki, *Terahertz Imaging System Using High-$T_c$ Superconducting Oscillation Devices*, J Appl Phys **111**, 123111 (2012).

[25] T. Kashiwagi et al., *Computed Tomography Image Using Sub-Terahertz Waves Generated from a High-$T_c$ Superconducting Intrinsic Josephson Junction Oscillator*, Appl Phys Lett **104**, 082603 (2014).

[26] T. Kashiwagi et al., *Reflection Type of Terahertz Imaging System Using a High-$T_c$ Superconducting Oscillator*, Appl Phys Lett **104**, 022601 (2014).

[27] Y. Saiwai, T. Kashiwagi, K. Nakade, M. Tsujimoto, H. Minami, R. A. Klemm, and K. Kadowaki, *Liquid Helium-Free High-$T_c$ Superconducting Terahertz Emission System and Its Applications*, Jpn J Appl Phys **59**, 105004 (2020).

[28] A. P. Zhuravel, A. V Ustinov, D. Abraimov, and S. M. Anlage, *Imaging Local Sources of Intermodulation in Superconducting Microwave Devices*, IEEE Transactions on Applied Superconductivity **13**, 340 (2003).

[29] A. P. Zhuravel, S. M. Anlage, and A. V Ustinov, *Imaging of Microscopic Sources of Resistive and Reactive Nonlinearities in Superconducting Microwave Devices*, IEEE Transactions on Applied Superconductivity **17**, 902 (2007).

[30] M. Tsujimoto, T. Kashiwagi, H. Minami, and K. Kadowaki, *Broadly Tunable CW Terahertz Sources Using Intrinsic Josephson Junction Stacks in High-Temperature Superconductors*, Terahertz Spectroscopy - A Cutting Edge Technology, Chapter 10, pp. 192 (2017), IntechOpen, London, UK.

[31] M. Tsujimoto, K. Yamaki, K. Deguchi, T. Yamamoto, T. Kashiwagi, H. Minami, M. Tachiki, K. Kadowaki, and R. A. Klemm, *Geometrical Resonance Conditions for THz Radiation from the Intrinsic Josephson Junctions in $Bi_2Sr_2CaCu_2O_{8+\delta}$*, Phys Rev Lett **105**, 037005 (2010).

[32] K. Delfanazari et al., *Tunable Terahertz Emission from the Intrinsic Josephson Junctions in Acute Isosceles Triangular $Bi_2Sr_2CaCu_2O_{8+\delta}$ Mesas*, Opt Express **21**, 2171 (2013).

[33] K. Delfanazari et al., *Study of Coherent and Continuous Terahertz Wave Emission in Equilateral Triangular Mesas of Superconducting $Bi_2Sr_2CaCu_2O_{8+\delta}$ Intrinsic Josephson Junctions*, Physica C Supercond **491**, 16 (2013).

[34] K. Delfanazari, H. Asai, M. Tsujimoto, T. Kashiwagi, T. Kitamura, T. Yamamoto, W. Wilson, R. A. Klemm, T. Hattori, and K. Kadowaki, *Effect of Bias Electrode Position on Terahertz Radiation from Pentagonal Mesas of Superconducting $Bi_2Sr_2CaCu_2O_{8+\delta}$*, IEEE Transactions on Terahertz Science and Technology **5**, 505 (2015).

[35] M. Tsujimoto, I. Kakeya, T. Kashiwagi, H. Minami, and K. Kadowaki, *Cavity Mode Identification for Coherent Terahertz Emission from High-$T_c$ Superconductors*, Opt Express **24**, 4591 (2016).

[36] T. Kashiwagi et al., *Improved Excitation Mode Selectivity of High-$T_c$ superconducting Terahertz Emitters*, J Appl Phys **124**, 033901 (2018).

[37] M. Tsujimoto et al., *Design and Characterization of Microstrip Patch Antennas for High-$T_c$ Superconducting Terahertz Emitters*, Opt Express **29**, 16980 (2021).

[38] G. Kuwano et al., *Experimental Validation of a Microstrip Antenna Model for High-$T_c$ Superconducting Terahertz Emitters*, J Appl Phys **129**, 223905 (2021).

[39] N. V. Chernomyrdin, A. O. Schadko, S. P. Lebedev, V. L. Tolstoguzov, V. N. Kurlov, I. V. Reshetov, I. E. Spektor, M. Skorobogatiy, S. O. Yurchenko, and K. I. Zaytsev, *Solid Immersion Terahertz Imaging with Sub-Wavelength Resolution*, Appl Phys Lett **110**, 221109 (2017).

[40] J. F. Federici, D. Gary, B. Schulkin, F. Huang, H. Altan, R. Barat, and D. Zimdars, *Terahertz Imaging Using an Interferometric Array*, Appl Phys Lett **83**, 2477 (2003).

[41] J.-Y. Lu, C.-C. Kuo, C.-M. Chiu, H.-W. Chen, Y.-J. Hwang, C.-L. Pan, and C.-K. Sun, *THz Interferometric Imaging Using Subwavelength Plastic Fiber Based THz Endoscopes*, Opt Express **16**, 2494 (2008).

[42] A. Zahid, H. T. Abbas, A. Ren, A. Alomainy, M. A. Imran, and Q. H. Abbasi, *Application of Terahertz Sensing at Nano-Scale for Precision Agriculture*, in *Wireless Automation as an Enabler for the Next Industrial Revolution* (Wiley, 2020), pp. 241–257.

[43] M. Gezimati and G. Singh, *Terahertz Imaging and Sensing for Healthcare: Current Status and Future Perspectives*, IEEE Access **11**, 18590 (2023).

[44] X. B. Huang, D. B. Hou, P. J. Huang, Y. H. Ma, X. Li, and G. X. Zhang, *The Meat Freshness Detection Based on Terahertz Wave*, in *Selected Papers of the Photoelectronic Technology Committee Conferences*, edited by S. Liu, S. Zhuang, M. I. Petelin, and L. Xiang, Vol. 9795 (SPIE, 2015), p. 97953D.

[45] T. Löffler, T. Bauer, K. J. Siebert, H. G. Roskos, A. Fitzgerald, and S. Czasch, *Terahertz Dark-Field Imaging of Biomedical Tissue*, Opt Express **9**, 616 (2001).

[46] A. J. Fitzgerald, E. Berry, N. N. Zinovev, G. C. Walker, M. A. Smith, and J. M. Chamberlain, *An Introduction to Medical Imaging with Coherent Terahertz Frequency Radiation*, Phys Med Biol **47**, R67 (2002).